\documentclass[11pt]{article}
\usepackage{moriond,epsfig}
\usepackage[latin1]{inputenc}
\usepackage[OT1]{fontenc}

\def\be{\begin{equation}}
\def\ee{\end{equation}}
\def\bea{\begin{eqnarray}}
\def\eea{\end{eqnarray}}

\newcommand{\ca}{a}
\newcommand{\cb}{b}

\newcommand{\ogpre}{{\Omega_{\gamma_01}}}

\newcommand{\oaf}{{\Omega_{\ca1}}}

\newcommand{\obf}{{\Omega_{\cb1}}}

\newcommand{\ogf}{{\Omega_{\gamma_11}}}

\newcommand{\obs}{{\Omega_{\cb2}}}

\newcommand{\ogfs}{{\Omega_{\gamma_12}}}

\newcommand{\faf}{f_{\ca1}}
\newcommand{\fbf}{f_{\cb1}}

\newcommand{\fbs}{f_{\cb2}}

\newcommand{\fafexpr}{{3\oaf/\left[4\ogpre + 3\oaf + 3\obf\right]}}
\newcommand{\fbfexpr}{{3\obf/\left[4\ogpre + 3\oaf + 3\obf\right]}}
\newcommand{\fbsexpr}{{3\obs/\left[4\ogfs + 3\obs\right]}}

\newcommand{\ra}{r_{\ca}}
\newcommand{\rb}{r_{\cb}}
\newcommand{\fNL}{f_{\rm{NL}}}

\newcommand{\fNLsingle}{\fNL^{\rm single}}

\newcommand{\zgpre}{\zeta_{\gamma_0}}

\newcommand{\zgf}{\zeta_{\gamma_1}}

\newcommand{\zgs}{\zeta_{\gamma_2}}

\newcommand{\za}{\zeta_{\ca}}

\newcommand{\zb}{\zeta_{\cb}}

\newcommand{\zf}{\zeta_{1}}

\newcommand{\zs}{\zeta_{2}}

\newcommand{\commentj}[1]{}

\begin{document}
%\vspace*{4cm}
\title{PRIMORDIAL NON-GAUSSIANITY FROM MULTIPLE CURVATON DECAYS}

\author{ JUSSI VALIVIITA, HOOSHYAR ASSADULLAHI and DAVID WANDS }

\address{Institute of Cosmology and Gravitation, University of
  Portsmouth, Portsmouth PO1 2EG, UK}

\maketitle\abstracts{We study a model where two scalar fields, that are subdominant during inflation, decay into radiation some time after inflation has ended but before primordial nucleosynthesis. Perturbations of these two curvaton fields can be responsible for the primordial curvature perturbation.
We write down the full non-linear equations that relate the primordial
perturbation to the curvaton perturbations on large scales, and solve them 
in a sudden-decay approximation.
We calculate the power spectrum of the primordial perturbation, and finally go to second order to find the non-linearity parameter, $\fNL$. {\em Not surprisingly}, we find large positive values of $\fNL$ if the energy densities of the curvatons are sub-dominant when they decay, as in the single curvaton case. But we also find {\em a novel effect}, which can be present only in multi-curvaton models: {\em $\fNL$ becomes large even if the curvatons dominate the total energy density} in the case when the inhomogeneous radiation produced by the first curvaton decay is diluted by the decay of a second nearly homogeneous curvaton. The minimum value $\min(\fNL)=-5/4$ which we find is the same as in the single-curvaton case.
Using (non-)Gaussianity observations, Planck can be able to distinguish between single-field inflation and curvaton model. Hence it is important to derive theoretical predictions for curvaton model. From particle physics point of view it is more natural to assume multiple scalar fields (rather than just one ``curvaton'' in addition to inflaton). Our work updates the theoretical predictions of curvaton model to this case.
}

\section{Introduction to curvaton model}

Theories beyond the standard model often contain a large number of scalar
fields in addition to the standard-model fields. In the very early universe it
is natural to expect the initial values of these fields to be displaced from
the minimum of their potential. If they are displaced by more than the Planck
scale, then they can drive a period of inflation. 
%But if they are displaced from their minimum by less than the Planck scale 
Otherwise they will oscillate
about the minimum of their potential once the Hubble rate, $H$, drops below
their effective mass. An oscillating massive field has the equation of state
(averaged over several oscillations) of a pressureless fluid. Thus the energy
density of a weakly interacting massive field tends to grow relative to
radiation in the early universe. Such fields must therefore decay before
the primordial nucleosynthesis era to avoid spoiling the standard, successful
hot big bang model. If the energy density of a late-decaying scalar is
non-negligible when it decays, then any inhomogeneity in its energy density
will be transfered to the primordial radiation.\cite{Mollerach:1989hu} %,Linde:1996gt}.
Therefore, the observed perturbations in the Universe can solely result from
these inhomogeneities and the perturbations in the inflaton decay
products (i.e., pre-existing radiation) can be negligible. This is the curvaton
scenario for the origin of structure.\cite{Enqvist:2001zp} %,Lyth:2001nq,Moroi:2001ct}.
We focus on this {\em pure} curvaton model.\footnote{{\em Before their decay},
the curvatons represent pure isocurvature degrees of freedom which are ruled
out by data.\cite{Enqvist:2000hp} Nevertheless, the pure curvaton model
generates {\em adiabatic primordial density perturbations}, if all the species are
in thermal equilibrium and the baryon asymmetry is generated after the curvatons
decay. 
However, if inflaton perturbation was not fully
negligible and the curvaton(s) decayed into radiation and cold dark
mater (CDM), then a correlated mixture of adiabatic and CDM isocurvature primordial perturbations
would follow. These possibilities have been studied extensively, e.g.,
in Refs.~\cite{Malik:2002jb}.   %\cite{Ferrer:2004nv}. 
It should be noted that an admixture of subdominant
isocurvature (up to 5-10\%) and predominant adiabatic
primordial perturbation can fit well~\cite{KurkiSuonio:2004mn} the current data.
%microwave background
%and large scale structure observations 
%though in the WMAP 5-year data
%the non-adiabatic feature of the second acoustic peak has disappeared
%with the new beam calibration.
} 

\section{Definitions: primordial perturbation, power spectrum, and $\fNL$}

The primordial density perturbation can be described in terms of the
non-linearly perturbed expansion on uniform-density 
hypersurfaces~\cite{Lyth:2004gb}
\begin{equation}
 \label{eqn:zetanl}
\zeta (t,{\bf x}) = \delta N (t,{\bf x}) +
 \frac13 \int_{\bar\rho(t)}^{\rho(t,{\bf x})}
 \frac{d\tilde\rho}{\tilde\rho+\tilde{p}} \,,
\end{equation}
where $N=\int Hdt$ is the integrated local expansion, $\tilde\rho$ the
local density and $\tilde{p}$ the local pressure, and $\bar\rho$
is the homogeneous density in the background model.
We will expand the curvature perturbation at each order $(n)$ as
%\begin{equation}
$\zeta(t,{\bf x}) = \sum_{n=1}^{\infty}\frac{1}{n!}\zeta_{(n)}(t, {\bf x})\,,$
%\label{eqn:expansion}
%\end{equation}
where we assume that the first-order perturbation, $\zeta_{(1)}$, is Gaussian as it is
proportional to the initial Gaussian field perturbations.
Higher-order terms describe the non-Gaussianity of the full non-linear $\zeta$.

Working in terms of the Fourier transform of $\zeta$, we define the primordial power spectrum as
%\begin{equation}
$\langle \zeta({\bf k_1}) \zeta({\bf k_2}) \rangle = (2\pi)^3
P_\zeta(k_1) \delta^3({\bf k_1}+{\bf k_2}) \,.$
%\end{equation}
The average power per logarithmic interval in Fourier space is given by
%\begin{equation}
 ${\cal P}_\zeta(k) = \frac{4\pi k^3}{(2\pi)^3} P_\zeta(k) \,,$
%\end{equation}
and is roughly independent of wavenumber $k$. The primordial bispectrum is given by
%\begin{equation}
% \label{defB}
$\langle \zeta({\bf k_1})\zeta({\bf k_2})
\zeta({\bf k_3}) \rangle
 =  (2\pi)^3 B(k_1,k_2,k_3) \delta^3({\bf k_1}+{\bf
k_2}+{\bf k_3}) \,.$
%\end{equation}
%
%The bispectrum vanishes for a purely Gaussian distribution, and hence is non-zero only at fourth and higher-order.
%
The amplitude of the bispectrum relative to the power spectrum is commonly parameterized in terms of the non-linearity parameter, $\fNL$, defined such that
%\begin{eqnarray}
$B(k_1,k_2,k_3) \!\!  =  \!\! (6/5)\fNL \left[ P(k_1) P(k_2) + 2\,{\rm perms} \right]\,.$
%\label{fnl}
%\end{eqnarray}
%Higher order statistics, like trispectrum
%(see e.g. \cite{Sasaki:2006kq,Byrnes:2006vq,Byrnes:2007tm}) or full non-linear probability density
%function of the primordial $\zeta$  \cite{Sasaki:2006kq}, can give also valuable information
%on non-Gaussianity in the curvaton model, but in this paper we consider the bispectrum only.

As an example we will consider the primordial curvature perturbation
produced by the decay of {\em two scalar fields} $a$ and $b$.
Without loss of generality, we assume that the curvaton $a$ decays first when $H=\Gamma_a$ followed by the decay of the curvaton $b$ when $H=\Gamma_b$, where $\Gamma_b<\Gamma_a$.

%
%Small-scale (sub-Hubble) vacuum fluctuations of any light scalar field are stretched by the expansion to large (super-Hubble) scales during a period of inflation in the early universe.
%
%If the curvaton fields are weakly-coupled, massive scalar fields whose masses are less than the
%Hubble scale, $H_\ast \gg m$, during inflation, then they 
During inflation the curvatons acquire an almost scale-invariant spectrum
of perturbations on super-Hubble scales,
%\begin{equation}
$ {\cal P}_{\delta\ca_{\ast}} = {\cal P}_{\delta\cb_{\ast}} = H_\ast^2/(2\pi)^2,$
%\end{equation}
where $H_\ast$ is the Hubble rate at Hubble exit.
%In Refs.~\cite{Seery:2005gb,Seery:2006js,Seery:2006vu,Yokoyama:2007uu,Sasaki:2007ay}
%it has been shown that in general,
%if slow-roll conditions are satisfied, the non-Gaussianity
%of field perturbations at Hubble exit is small
%(at least for the three-point and four-point correlators). Consequently,
%in what follows,
We assume that the field perturbations at Hubble exit have
Gaussian and independent distributions, consistent with weakly-coupled isocurvature
fluctuations of light scalar fields.
% \cite{Enqvist:2004bk,Vaihkonen:2005hk}.
When 
%(some time after the end of inflation)
the Hubble rate drops below
the mass of each curvaton field, the field begins to oscillate.
The local value of each curvaton can evolve between Hubble-exit during inflation
and the beginning of the field oscillations.
We parameterize this evolution by functions $g_{\ca}$ and $g_{\cb}$, but we
assume the two fields
remain decoupled so that their perturbations remain uncorrelated. Thus at the
beginning of curvaton oscillations, the curvaton fields have values
%\begin{eqnarray}
%\label{defga}
$\ca_{\rm osc}  =  g_{\ca}(\ca_\ast)$ and
%\label{defgb}
$\cb_{\rm osc}  =  g_{\cb}(\cb_\ast)\,.$
%\end{eqnarray}

We can define non-linear
perturbation $\za$  and $\zb$ for each curvaton
analogous to the total
perturbation (\ref{eqn:zetanl}).
They become constant on scales larger than the Hubble scale once
each curvaton starts oscillating, and while we can neglect energy transfer to the radiation,
$\Gamma_\ca<H$.
We find that the power spectra of $\za$ and $\zb$ when the curvatons start to oscillate
are related by
%\begin{equation}
$P_{\zb} = \beta^2 P_{\za}\,,$
%\label{eqn:PbvsPa}
%\end{equation}
where
$\beta = \frac{g'_{\cb}/g_{\cb}}{g'_{\ca}/g_{\ca}}$
%\end{equation}
with $g'_{\cb}\equiv \partial g_{\cb} / \partial\cb_{\ast}$
and similar for $g'_{\ca}$.
In this talk, we assume linear evolution between Hubble exit and the beginning
of curvaton oscillations, so that $\beta$ reduces to the ratio
of the background curvaton field values at Hubble exit,
%\begin{equation}
$\beta = \ca_{\ast}/\cb_{\ast}\,.$
%\end{equation}

%Now we need to calculate what kind of total (radiation)
%curvature perturbation follows form the decay of $\ca$ and
%$\cb$.

\section{Full non-linear equations}

We will estimate the primordial density perturbation produced by the
decay of two curvaton fields some time after inflation has ended
using the sudden-decay approximation,
generalizing the non-linear analysis of Ref.~\cite{Sasaki:2006kq} to
the case of two curvatons. Before the decays we have three
non-interacting fluids with barotropic equations of state and hence three curvature
perturbations (\ref{eqn:zetanl}) which are constant on large scales
\begin{equation}\textstyle
 \zeta_\gamma = \delta N + \frac14 \ln \left(\frac{\rho_{\gamma}}{\bar\rho_{\gamma}} \right) \,,%\label{eqn:zgfullnl}
\quad\quad
 \za = \delta N + \frac13 \ln \left( \frac{\rho_a}{\bar\rho_a} \right) \,,%\label{eqn:zafullnl}
\quad\quad
 \zb = \delta N + \frac13 \ln \left( \frac{\rho_b}{\bar\rho_b} \right) \,\label{eqn:zbfullnl}.
\end{equation}

On the spatial hypersurface where $H=\Gamma_a$, there is an abrupt
jump in the overall equation of state due to the sudden decay of the
first curvaton into radiation, but the total energy density is
continuous
\begin{equation}
\rho_{\gamma_11} + \rho_{\cb1} = \rho_{\gamma_01} + \rho_{\ca1} + \rho_{\cb1}\,.
\label{eqn:totrhoat1}
\end{equation}
Here $\rho_{\gamma_11}$ is the radiation energy density immediately after the
first curvaton decay, $\rho_{\cb1}$ is density of the second curvaton at 
%the time of 
the first curvaton decay, and $\rho_{\gamma_01}$ and $\rho_{\ca1}$ are densities of
pre-existing radiation the first curvaton just before the first decay where
$\ca$ is converted into $\gamma_1$.
% and $\rho_{\ca1}$ is
%the density of the first curvaton just before the first decay when it is converted
%to radiation.
%
Note that the decay hypersurface is a uniform-density hypersurface and
thus, from Eq.~(\ref{eqn:zetanl}), the perturbed expansion on this hypersurface
is $\delta N=\zf$, where $\zeta_1$ denotes the total curvature perturbation
at the first-decay hypersurface. Hence employing Eq.~(\ref{eqn:zbfullnl})
we have on this hypersurface
 \begin{equation}
 \rho_{\gamma_01} =  \bar\rho_{\gamma_01} e^{4(\zgpre-\zf)} \,,
\quad \
 \rho_{\gamma_11}  =  \bar\rho_{\gamma_11} e^{4(\zgf-\zf)} \,,
\quad \
 \rho_{\ca1}  =  \bar\rho_{\ca1} e^{3(\za-\zf)} \,,
\quad \
 \rho_{\cb1}  =  \bar\rho_{\cb1} e^{3(\zb-\zf)} \,.\label{eqn:rhob1}
\end{equation}
%where we employed Eq.~(\ref{eqn:zbfullnl}).

On a uniform-density hypersurface the total energy density
is homogeneous. Therefore an infinitesimal time before the first decay
we have $\rho_{\gamma_01}+\rho_{\ca1}+\rho_{\cb1} = \bar\rho_1.$
Substituting here Eqs. (\ref{eqn:rhob1}), and dividing by the total energy density
$\bar\rho_1$ we end up with (assuming flat $\Omega=1$ universe)
\begin{equation}
  \ogpre e^{4(\zgpre - \zf)} + \oaf e^{3(\za - \zf)} + \obf e^{3(\zb - \zf)} =
  1\,,
\label{eqn:nonlinf1}
\end{equation}
where $\ogpre = \rho_{\gamma_01}/\bar\rho_1$, $\oaf=\rho_{\ca1}/\bar\rho_1$,
and $\obf=\rho_{\cb1}/\bar\rho_1$ are the energy density parameters
of the pre-existing radiation, the first curvaton, and the second curvaton at
the first curvaton decay, respectively.
An infinitesimal time after the first curvaton has
decayed into radiation
we have
\begin{equation}
  \ogf e^{4(\zgf - \zf)} + \obf e^{3(\zb - \zf)} = 1\,.
\label{eqn:nonlinf2}
\end{equation}
The total curvature perturbation $\zf$ is the perturbation on
the decay hypersurface and thus is continuous between the two phases.
However, the energy density of radiation changes abruptly at the decay
time and the radiation curvature perturbation is discontinuous,
$\zgpre\neq\zgf$. Eqs. (\ref{eqn:nonlinf1}) and (\ref{eqn:nonlinf2})
can be used also for the second curvaton decay with the obvious
substitutions  $\obf \rightarrow 0$, index $a \rightarrow b$, $1 \rightarrow 2$, $0 \rightarrow 1$.
Results for single curvaton follow by setting the
density parameters of $\cb$ zero. In  Ref.~\cite{Sasaki:2006kq}
we have checked in single-curvaton case that a fully nonlinear numerical
solution for a gradual decay agrees well with the analytical solution
obtainable from (\ref{eqn:nonlinf1}) and (\ref{eqn:nonlinf2}). 

If one is interested in perturbative results, one can Taylor expand the
exponential functions, $e^x = 1 + x + x^2/2 + \ldots$, and substitute
the expansions of the curvature perturbations up to any desired order, and
then solve order by order for the
final curvature perturbation $\zs=\zgs$ as a function of $\zgpre$, $\za$, and $\zb$.
Indeed in the {\em pure} curvaton model the pre-existing radiation
from inflaton decay is homogeneous $\zgpre=0$, which we assume from now on.  

\section{Results: Primordial power spectrum and $\fNL$}

The first order curvature transfer efficiency parameters~\cite{Assadullahi:2007uw}
can be given in terms of
three ratios which specify the relative densities
at the first decay
$\faf = \fafexpr$, 
$\fbf = \fbfexpr$
and at the second decay
$\fbs = \fbsexpr$.

Solving Eqs.~(\ref{eqn:nonlinf1}) and (\ref{eqn:nonlinf2}) up to first order,
we obtain the primordial power spectrum after the
second curvaton decay~\cite{Assadullahi:2007uw}:
%\begin{equation}
$P_{\zeta} = \left[\ra^2+\beta^2\rb^2\right]P_{\za}\,,$
%\end{equation}
where $\beta^2$ gives the ratio between the initial power
in the curvaton $\ca$ and that in the curvaton $\cb$.  
Here the total first order curvaton perturbation transfer efficiencies
are~\footnote{Note that our result differs from that presented recently
by Choi and Gong \cite{Choi:2007fya} due to the presence of the extra 
term $\fbf$ which they (implicitly) assumed zero.
The term  $\fbf$ arises due to the difference between the uniform total density hypersurface
and the uniform radiation density hypersurface when curvaton $\ca$ decays,
if the density of the curvaton $\cb$ is not negligible. When $\obf=0$ we
we recover the  simpler result~\cite{Choi:2007fya} $\ra=(1-\fbs)\faf$ and $\rb=\fbs$.}
$\ra = \left[(1\!-\!\fbs)(3\!+\!\faf)\faf\right]/\left[3(1\!-\!\fbf)+\faf\right]$ and
$\rb = \left[(1\!-\!\fbf)\fbs(3\!+\!\faf)+\fbf\faf\right]/\left[3(1\!-\!\fbf)+\faf\right]$.
Already this first order result is much more cumbersome than in the single
curvaton 
case.\footnote{If the other curvaton, say $\cb$, is absent, then we recover the standard
single-curvaton first-order result $P_{\zeta} = \ra P_{\za}$ with 
$\ra=\faf=3\oaf/\left[4\ogpre + 3\oaf \right] = 3\oaf/\left[4 - \oaf \right]$.}

\begin{figure}[t]
\begin{center}
\epsfig{figure=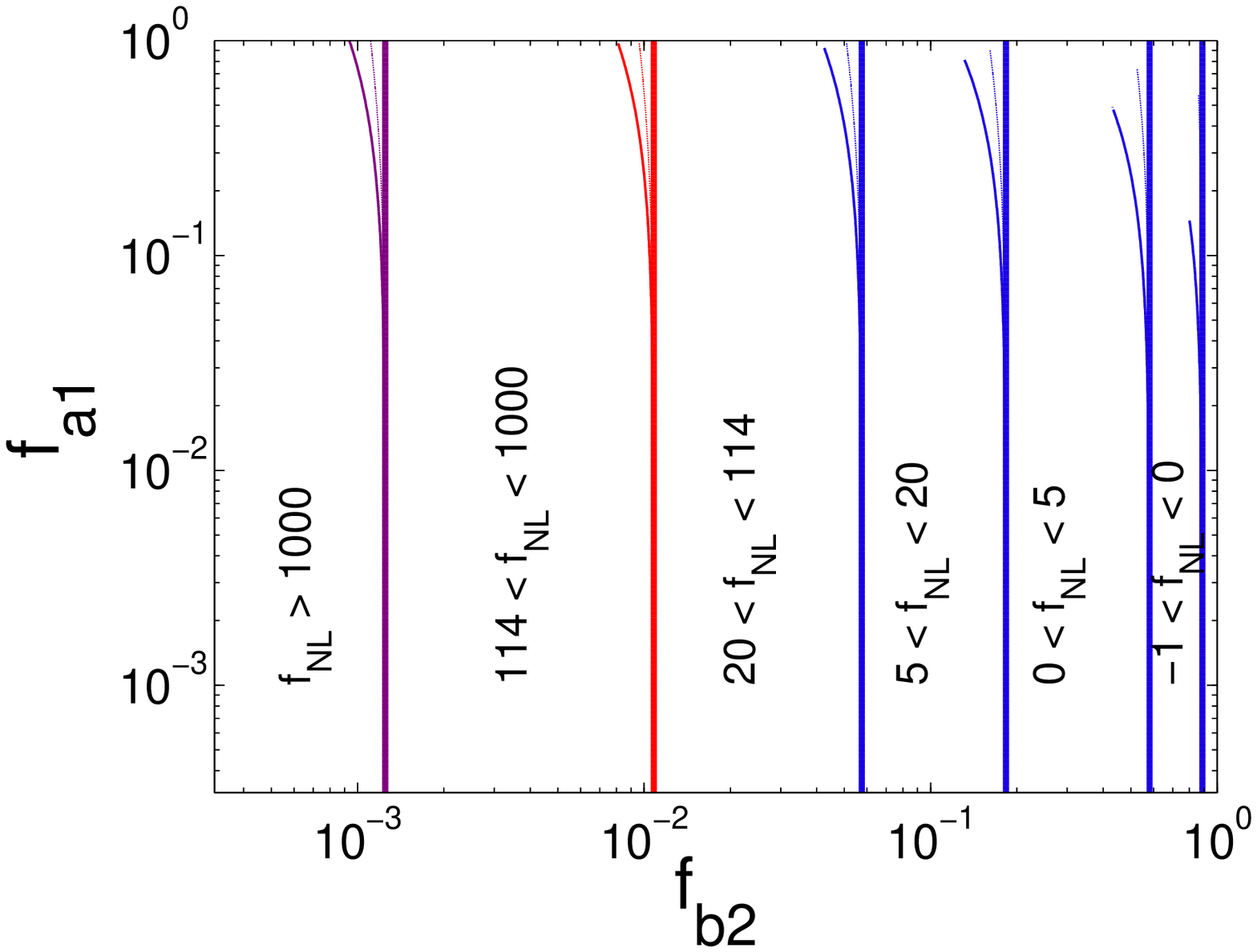,width=0.327\textwidth}
\epsfig{figure=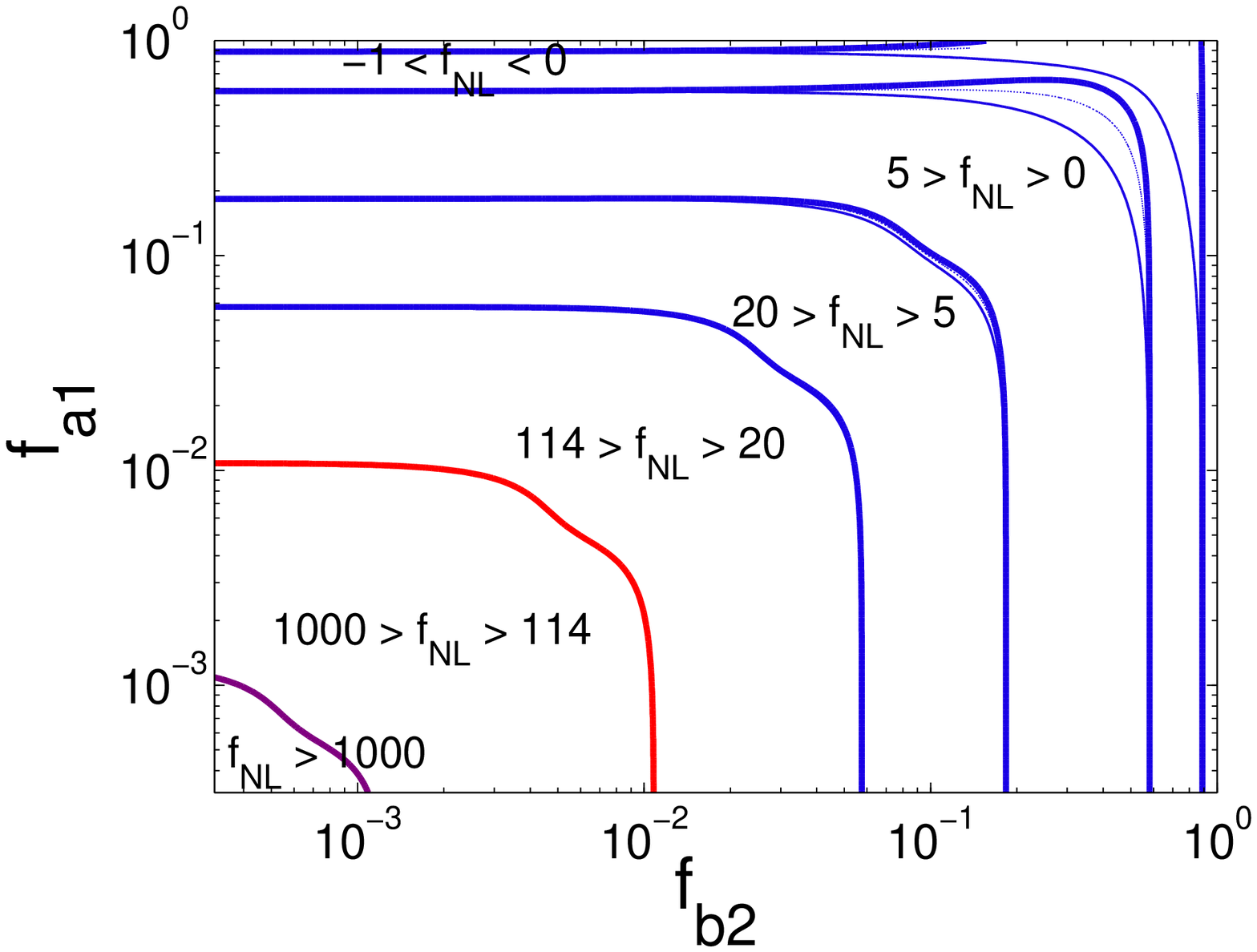,width=0.327\textwidth}
\epsfig{figure=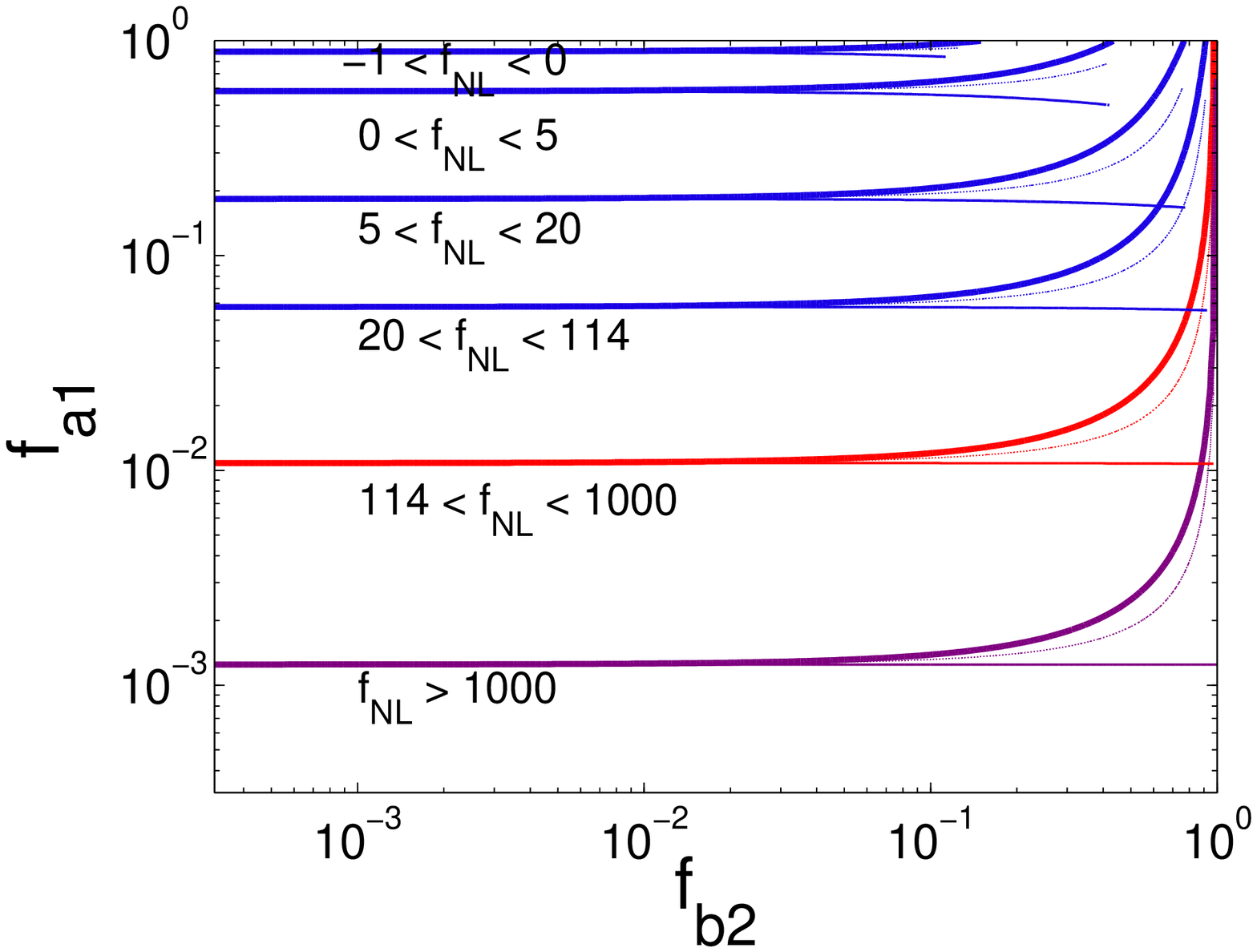,width=0.327\textwidth}
\end{center}
\caption{$\fNL(\fbs,\faf)$ for $\beta=\infty$ (left panel), $\beta=1$ (middle
  panel), and $\beta=0$ (right panel). The energy density ratio of the first
  curvaton at the first decay is set to $\fbf=0$ (\emph{thick
solid lines}) or $\fbf=\fbs/2$
  (\emph{dotted lines}) or $\fbf=(1+\faf/3)\fbs$ (\emph{thin solid lines})
  which corresponds to the simultaneous decay of $\ca$ and $\cb$.
  Contours of equal $\fNL$ are shown, from left to right (or from bottom to
  top),
  for $\fNL=$ $1000$, $114$, $20$,
  $5$, $0$, and $-1$.\hspace{\fill}
\label{fig:Valiviita}}
\end{figure}

In Ref.~\cite{Assadullahi:2007uw} we solve  Eqs.~(\ref{eqn:nonlinf1}) and (\ref{eqn:nonlinf2})
step by step up to second order for the first and second decay, and finally
give an expression for the primordial non-linearity parameter $\fNL$ after the
second decay. It is a very complicated function of $\beta$, $\faf$, 
$\fbf$ and $\fbs$. In 3 panels in Fig.~\ref{fig:Valiviita} we show it for
$\beta= \infty, 0,\,1$
and for three choices of $\fbf$ (with different line styles). 
(a) $\beta \rightarrow \infty$: in this case the first curvaton
$\ca$ is almost homogeneous. Then the result is very close to
the standard single-curvaton result (in Fig.\ $\fbf=0$);
$\fNLsingle(f) = \frac{5}{4f} - \frac{5}{3} -\frac{5f}{6}$
with $f=\fbs$. However, if the first curvaton density at the
first decay is non-negligible, it modifies the result slightly.
Large $\fNL$ is obtained if $\cb$ is highly subdominant
($\fbs\ll 1$) at its decay time.
(b) $\beta=0$: both curvatons carry equal
amount of perturbation. The result is what one would expect from
single-curvaton studies.  Large $\fNL$ is obtained if and only if both
$\ca$ and $\cb$ are highly subdominant ($\faf\ll 1, \fbs\ll 1$)
at their decay time. (c) $\beta=0$: the
second curvaton is homogeneous. In this case we can obtain
a large  $\fNL$ even if both the curvatons are
dominant ($\faf\sim 1, \fbs\sim 1$, upper right corner in Fig.)
at their decay time. The inhomogeneous radiation produced by the first
curvaton decay is diluted by the decay of the second homogeneous
curvaton. We expect this result to hold also for more
than two curvatons if the last decaying curvaton is almost
homogeneous.

In all cases we find $\fNL\geq-5/4$, which seems to be a robust
lower bound in single and multi-field curvaton models in which the
curvaton field perturbations are themselves Gaussian.
% which should be a good approximation for a weakly interacting field.

\section*{Acknowledgments}
JV is supported by STFC and the Academy of Finland grants 120181\&125688.

%%%%%%%%%%%%%%%%%%%%%%%%%%%%%%%%%%%%%%%%%%%%%%%%%%%%%

\end{document}